\documentclass[preprint,amsmath,amssymb,prd]{revtex4-2}

\usepackage{hyperref}
\usepackage{dsfont}
\usepackage{stmaryrd}
\usepackage{graphicx}

\begin{document}

\title{Perturbative Algebraic Approach to Interacting  Quantum Fields on the DFR Quantum Spacetime}

\author{J.F. L\'opez}
\email[]{jf.lopezr@uniandes.edu.co}
\affiliation{Departamento de F\'{i}sica, Universidad de los Andes,  A.A. 4976-12340, Bogot\'a, Colombia}

\author{A.F. Reyes-Lega}
\email[]{anreyes@uniandes.edu.co}
\affiliation{Departamento de F\'{i}sica, Universidad de los Andes,  A.A. 4976-12340, Bogot\'a, Colombia}

\begin{abstract}
An approach to renormalization of scalar fields on the Doplicher-Fredenhagen-Roberts (DFR) quantum spacetime is presented. The effective non-local theory obtained through the use of states of optimal localization for the quantum spacetime is reformulated in the language of (perturbative) Algebraic Quantum Field Theory. The structure of the singularities associated to the non-local kernel that codifies the effects of noncommutativity is analyzed using the tools of microlocal analysis. 
\end{abstract}

\maketitle
\newpage
\tableofcontents

\section{Introduction}	
One of the most difficult problems in theoretical physics has been, for many decades, the formulation of a consistent quantum theory of gravity. Apart from the conceptual and mathematical difficulties, the lack of observational evidence for  the quantum nature of spacetime at the Planck scale has been a major obstacle. However, this state of affairs is likely to change in the forthcoming years, as the emergence of the multi-messenger paradigm seems to indicate~\cite{Meszaros2019,addazi2022quantum}.

Well established approaches to the problem (like, e.g.,  loop quantum gravity~\cite{Rovelli2008}, asymptotic safety~\cite{Niedermaier2006} or causal sets~\cite{Surya2019}) aim for a description of quantum gravity at a fundamental level. In a more phenomenological vein, there are other approaches that follow a ``bottom-up'' approach. That is the case with several models of quantum spacetimes which incorporate the idea of a fundamental minimal length scale, but still can be regarded as ``deformations'' of, say, Minkowski spacetime. The idea of  introducing a set of noncommutative coordinate operators describing a quantum spacetime goes back to Heisenberg and Snyder~\cite{Hossenfelder2013}. Much later, similar models were found as low-energy limits of string theories~\cite{seiberg1999string}. The study of quantum field theory on such quantum spacetimes has been a subject of notorious interest during the last two decades.
Since --by construction-- such models still bear resemblance to the physics of lower energy scales, they offer an interesting possibility for the exploration of quantum gravity signatures.

In 1994, Doplicher, Fredenhangen and Roberts (DFR) proposed a model for a quantum spacetime that retains all the symmetries of Minkwoski spacetime, but that also includes a parameter $\lambda_P$ (of the order of magnitude of Planck's length) which reflects the quantum nature of spacetime. Heuristically, the idea is to promote the coordinates of events in Minkowski spacetime to operators satisfying commutation relations that lead to uncertainty relations compatible with the \emph{Principle of Gravitational
Stability against localisation of events} \cite{doplicher1994spacetime}.

This principle basically says that events, regarded as points in a spacetime $\mathcal{M}$ that is  modeled as a smooth manifold, do not have  a physically coherent interpretation at small length scales (of the order of magnitude of Planck's length). 
The physical existence of a point $x\in\mathcal{M}$, in the setting of Algebraic QFT (AQFT), would require to define a field localized in a region of $\mathcal{M}$ arbitrarily close to $x$. From the Heisenberg uncertainty principle $\Delta E\Delta t\gtrsim \hbar/2$, localization in time increases uncertainty in energy density, giving place to creation and annihilation of a lot of particles (of the theory). If this energy density is localized in a region smaller than its Schwarszchild radius, it  would generate a trapped surface (e.g. a black hole) hiding the event at all (for an interesting discussion of this issue in the context of QFT on curved spacetimes with back reaction, see \cite{doplicher2013quantum}). The principle solves this problem by imposing  certain \emph{uncertainty relations} on the spacetime coordinates that forbid such extreme localization of events.

Mathematically, the model is formulated in terms of a $C^*$-algebra, which is to be interpreted as the ``algebra of functions'' on this quantum spacetime, much in the spirit of noncommutative geometry \cite{connes1995noncommutative}.
The possibility of constructing quantum fields over this quantum spacetime was already discussed in the original works \cite{doplicher1994spacetime,doplicher1995quantum}.

The theory of quantum fields over the DFR quantum spacetime has been the subject of intense study during the last two decades (see, e.g., \cite{bahns2003ultraviolet,bahns2005field}, or  \cite{bahns2015quantum}  for a  review). More recently,  techniques from perturbative Algebraic Quantum Field Theory (pAQFT) have been applied to models of quantum fields on the  DFR quantum spacetime \cite{doplicher2020perturbative}.
One of the main issues with standard interacting QFT is the  occurrence of (ultraviolet) divergent integrals order by order in the loop expansion in perturbation theory \cite{peskin2018introduction}.
Mathematically, it is possible to trace the origin of these divergences in Fourier space back to ill-defined products of distributions in coordinate space. Epstein-Glaser renormalization \cite{epstein1973role} is the scheme which solves the problem of divergences via the extension of distributions satisfying certain conditions in a completely finite way.
pAQFT \cite{dutsch2019classical,rejzner2018perturbative} has been recently developed in order to combine the formal understanding of Algebraic (or axiomatic) QFT with the perturbative methods. Its application to the study of quantum fields on the DFR quantum spacetime is an important addition to the already extensive body of work where
 other approaches to QFT have been extended to  this non-commutative Minkowski spacetime. Just to name a few, there is one approach  inspired in the normal order by point-splitting \cite{doplicher2013quantum}, another one based on the  Yang-Feldman equation \cite{doplicher2020quantum}.  The point-splitting version has been written in the pAQFT framework \cite{doplicher2020perturbative,bahns2004perturbative}. This model has been proved to be free of UV divergences and the existence of its adiabatic limit has been established.

 The different approaches to implement QFT over the noncommutative quantum spacetime, focused on defining the interacting normally ordered action functional, lead to inequivalent non-local effective models with very different singularity structures. 
 Nevertheless,  when the commutative limit is taken (by taking the limit $\lambda_P\rightarrow 0$), correlation functions computed using those different approaches give the standard results for QFT on Minkowski space.

 Therefore, even in the UV finite approach by point-splitting, some renormalization procedure will be necessary in order to compare with the commutative, low-energy regime. Deviations from the commutative approximation, in a  bottom-up approach, could give information about the (still  unknown) quantum gravity theory, or its effective low-energy theory.

In the present paper, we have proposed a map that assigns to any \emph{local} interaction for a scalar field on the DFR Quantum Spacetime a \emph{non-local} effective interacting quantum field theory in the framework of pAQFT \cite{dutsch2019classical}, based on the original Moyal type product of fields by Doplicher, Fredenhagen and Roberts \cite{doplicher1995quantum}. Moreover, the singularity structure of this new effective theory is described in terms of the wavefront set of some distribution \cite{hormander2015analysis}.

Let us finish this introduction with a description of the remaining sections of the paper. In section \ref{sec:2}, the DFR Quantum Spacetime is presented following the main  results of \cite{doplicher1995quantum}. Section \ref{sec:3} is devoted to a quick review of the basics of interacting pAQFT. The next two sections contain the main results of this paper. In section \ref{sec:4} 
we obtain an effective non-local interacting Quantum Field model and express it in the language of pAQFT. The modified Feynman diagrams that are obtained for this model and their corresponding rules are also presented. In section \ref{sec:5} we dicuss some aspects about unitarity of the modified S matrix. In section \ref{sec:6} we present an analysis of the  microlocal properties of the non-local kernel previously obtained.
We finish with some final remarks in section \ref{sec:7}

\section{Quantum Space-Time}
\label{sec:2}

A Quantum Space-Time (QST) shall be understood as a noncommutative space in the context of noncommutative geometry \cite{connes1995noncommutative, gracia2013elements}. By this, functions over the QST are elements of a C*-algebra $\mathcal{E}$ generated by the unbounded affiliated selfadjoint elements $\{q^{\mu}\}_{\mu=0}^{3}$, the coordinate operators; which satisfy following the Space-Time Uncertainty Relations (STUR) \cite{doplicher1994spacetime}

\begin{align}
    \label{stur1}
    \Delta q^0\sum_{j=1}^{3}\Delta q^i&\geq\frac{\lambda_P^2}{2},\\
    \label{stur2}
    \sum_{1\leq j< k\leq 3}\Delta q^i\Delta q^j&\geq\frac{\lambda_P^2}{2},
\end{align}

where $\lambda_P$ is some constant with units of the order of the Planck length. Here, an affiliated element $q$ affiliated to $\mathcal{E}$ is taken in the sense of Woronowicz \cite{woronowicz1995c} as the C*-homomorphism

\begin{equation}
    \begin{matrix}
        C_0^{\infty}(\mathbb{R}) & \rightarrow & M(\mathcal{E})\\
        f & \mapsto & f(q),
    \end{matrix}
\end{equation}

where $M$ stands for the multiplier algebra \cite{busby1968double}. A state $\omega\in\mathcal{S}(\mathcal{E})$ is in the domain of $q$ if

\begin{equation}
    sup\left\{\omega(f(q))|f\in\ C_0(\mathbb{R}),\ f(x)<x^2\ \forall x\in\mathbb{R}\right\}<\infty,
\end{equation}

and the Space-Time Uncertainty Relations are assumed to hold in any state in the domain of all $q^{\mu}$, where $\Delta(q)=\sqrt{\omega(q^2)-\omega(q)^2}$. Doplicher, Fredenhagen and Roberts proposed \cite{doplicher1995quantum} an implementation of the \eqref{stur1} and \eqref{stur2} relations imposing commutation relations for the coordinate operators (this approach will be followed for the remaining of this section)

\begin{equation}
    \label{comm1}
    \left[q^{\mu},q^{\nu}\right]=i\lambda_P^2 Q^{\mu\nu},
\end{equation}

with $Q^{\mu\nu}$ selfadjoint elements which are assumed to be central elements (commuting with the coordinate generators)

\begin{equation}
    \label{comm2}
    \left[q^{\mu},Q^{\sigma\rho}\right]=0,
\end{equation}

and satisfying the Quantum Conditions

\begin{align}
    Q^{\mu\nu}Q_{\mu\nu} &= 0,\\
    \frac{1}{4}\left(\epsilon_{\mu\nu\sigma\rho}Q^{\mu\nu}Q^{\sigma\rho}\right)^2 &=1.
\end{align}

The C*-algebra generated by these operators is isomorphic to the trivial bundle of C*-algebras

\begin{equation}
    \mathcal{E}\cong C_0(\Sigma,\mathcal{K}),
\end{equation}

where $\mathcal{K}$ denotes the space of compact operators on a separable Hilbert space, and $\Sigma$ is the joint spectrum of the central operators $Q^{\mu\nu}$. Topologically, $\Sigma$ is the product of the tangent bundle of a sphere and a 2-point set

\begin{equation}
    \label{algebra}
    \Sigma\cong TS^2\times\{\pm\}\cong SL_2(\mathbb{C})/D.
\end{equation}

The second equality in \eqref{algebra} is obtained by considering the constant values $\sigma\in\Sigma$ of $Q$ on irreducible representations of $\mathcal{E}$ (the spectral values) as antisymmetric $4\times 4$ matrices with electric and magnetic components orthogonal and equal in norm. This is a homogeneous space with action of $SL_2(\mathbb{C})$ (the double covering of the proper Lorentz group) as a second rank tensor, and isotropy $D$ group the space of $2\times 2$ diagonal matrices. The only representations we shall be concerned with are those for which the Weyl-like operators $e^{i k_{\mu}q^{\mu}}$
 are well defined and are strongly continuous in each $k^{\mu}$. Those representations are called regular realizations. A faithful representation can be obtained via the Fourier transform. Using the central elements $Q$ as Casimir operators to label irreducible representations, a faithful representation has the direct integral form

 \begin{equation}
     \mathcal{E}\cong \int_{\oplus}\mathcal\pi_{\sigma}(\mathcal{E}),
 \end{equation}

 such that $\pi_{\sigma}(Q^{\mu\nu})=\sigma^{\mu\nu}$. For $F\in C_0(\Sigma,L^1(\mathbb{R}^4))$ such that $F(\sigma)(k)=\hat{f}(\sigma,k)$, it is possible to define the operator

 \begin{equation}
     \pi_{\sigma}(f(q))=\int \frac{dk}{(2\pi)^4}\hat{f}(\sigma,k)e^{ik_{\mu}\pi_{\sigma}(q^{\mu})};
 \end{equation}

 or in short notation

 \begin{equation}
     f(q)=\int \frac{dk}{(2\pi)^4}\hat{f}(Q,k)e^{ik_{\mu}q^{\mu}}.
 \end{equation}

 When restricted to a representation $\pi_{\sigma}$, the commutation relations \eqref{comm1}, \eqref{comm2} lead to the Moyal product

 \begin{equation}
     e^{ik^1_{\mu}q^{\mu}}*e^{ik^2_{\mu}q^{\mu}}=e^{i\frac{\lambda_P^2}{2}k^1_{\mu}\sigma^{\mu\nu}k^2_{\nu}}e^{i(k^1+k^2)_{\mu}q^{\mu}},
 \end{equation}

 which completely determines the algebra $\mathcal{E}$. Let us denote with $\mathcal{S}(\mathcal{E})$ the space of states of $\mathcal{E}$. Following, we will work with states $\omega\in\mathcal{S}(\mathcal{E})$ of the form

 \begin{equation}
    \label{state}
     \omega(f(q))=\int_{\Sigma} d\mu(\sigma)\int\frac{dk}{(2\pi)^4}\hat{f}(\sigma,k)\omega_{\sigma}(e^{ik_{\nu}\pi_{\sigma}(q^{\nu})}),
 \end{equation}

 where $\mu$ is a probability measure over $\Sigma$ and $\omega_{\sigma}$ is a continuous collection of states in each irreducible representation $\pi_{\sigma}$. An interesting example of a such a state in $\mathcal{E}$ is given by the optimally localized state around $x\in\mathbb{R}^4$. This state minimizes the uncertainty $\sum_{\nu}\Delta(q^{\mu})^2$ and it is given by a measure $\mu$ supported only on $\Sigma_1$ -the zero section of $\Sigma$- and

 \begin{equation}
    \label{invstate}
     \omega_{\sigma}(e^{ik_{\nu}\pi_{\sigma}(q^{\nu})})=exp\left[-\frac{\lambda_P^2}{2}\sum_{\nu}\left(k_{\nu}\right)^2\right]e^{ik_{\nu}x^{\nu}}.
 \end{equation}

 Clearly, this state is not Lorentz invariant, neither any state with a measure $\mu$ compactly supported in $\Sigma$. Moreover, measures with non-compact support allow arbitrary delocalized states (with localization measured by $\sum_{\mu}\Delta_{\omega} (q^{\mu})^2$). The simplest choice, which we will follow in this work, is to pick in each sphere the unique measure induced by rotational symmetry. Another state derived from \eqref{state}, takes into account the integral of the operator $f(q)$ in the whole euclidean space at a fixed time $t$. With the invariant measure at $\Sigma_1$ \eqref{invstate}

 \begin{equation}
     \int_{q^0=t}f(q)d^3q:=\int_{\Sigma_1}d\mu(\sigma)\int\frac{dk_0}{2\pi}\hat{f}(\sigma;k_0,\vec{0})e^{ik_0 t}.
 \end{equation}

 These type of expressions appear in the definition of the Hamiltonian at time $t$ in terms of some density (e.g. interactions), as it is presented below.

\section{Quantum Field Theory}
\label{sec:3}

In the following sections, the framework of perturbative Algebraic Field Theory \cite{dutsch2019classical,dutsch2003master} will be used. In this setting, the configuration space of \textit{off-shell} fields for the real scalar field is $\mathcal{C}:=C^{\infty}(\mathbb{R}^4)$, i.e. the equations of motion are not imposed at the very beginning. The fundamental fields are defined in the following way. For fixed $x\in\mathbb{R}^4$, we define $\phi(x)$ to be the evaluation map

 \begin{equation}
     \begin{split}
         \phi(x):  \mathcal{C} & \rightarrow  \mathbb{R}\\
          h & \mapsto  h(x).
     \end{split}
 \end{equation}

 Analogously, for $a\in\mathbb{N}^4$ $\partial^a\phi(x)$ is defined by

 \begin{equation}
     \begin{split}
         \partial^{a}\phi(x):  \mathcal{C} & \rightarrow  \mathbb{R}\\
          h & \mapsto  \partial^{a}h(x)=\frac{\partial^{|a|}h(x)}{\partial^{a_0}x^0\ldots\partial^{a_3}x^3},
     \end{split}
 \end{equation}

 where $|a|=a_0+...+a_3$. The set of fields (i.e. the observables of the theory), denoted by $\mathcal{F}$, are complex valued functions on the configuration $\mathcal{C}$ with all its functional derivatives well defined, and non-zero only up to finite order. That is, if $F\in\mathcal{F}$

%%%%%%%%%%%%%%%%%%%%%%%%%%%%%%%%%%%%%%%%%%%%
%%%%%%%%%%%%%%%%%%%%%%%%%%%%%%%%%%%%%%%%%%%%
%%%%%%%%%%%%%%%%%%%%%%%%%%%%%%%%%%%%%%%%%%%%
%%%%%%%%%%%%%%%%%%%%%%%%%%%%%%%%%%%%%%%%%%%%
% Cambia phi por varphi

 \begin{equation}
    \begin{split}
        F:\mathcal{C}&\rightarrow\mathbb{C}\\
        \phi&\mapsto F(\varphi)=f_0+\sum_{n=0}^{N}f_n(\varphi^{\otimes n}),
    \end{split}
 \end{equation}
 
%%%%%%%%%%%%%%%%%%%%%%%%%%%%%%%%%%%%%%%%%%%%
%%%%%%%%%%%%%%%%%%%%%%%%%%%%%%%%%%%%%%%%%%%%
%%%%%%%%%%%%%%%%%%%%%%%%%%%%%%%%%%%%%%%%%%%%
%%%%%%%%%%%%%%%%%%%%%%%%%%%%%%%%%%%%%%%%%%%% 

 where $f_0\in\mathbb{C}$ and $f_n$ are complex valued, compactly supported symmetric distributions in $\mathbb{R}^{4n}$; such that its wavefront set satisfies

 \begin{equation}
     WF(f_n) \cap \left(\overline{V_{+}^n}\cup\overline{V_{-}^n}\right)=\emptyset,
 \end{equation}

 where $\overline{V_{\pm}}$ are the closures of the forward/backward cones for a metric $\eta=diag(+,-,-,-)$ in Minkowski space-time given by

%%%%%%%%%%%%%%%%%%%%%%%%%%%%%%%%%%%%%%%%%%%%
%%%%%%%%%%%%%%%%%%%%%%%%%%%%%%%%%%%%%%%%%%%%
%%%%%%%%%%%%%%%%%%%%%%%%%%%%%%%%%%%%%%%%%%%%
%%%%%%%%%%%%%%%%%%%%%%%%%%%%%%%%%%%%%%%%%%%%
% Se incluyeron los exponentes n en V_+ y V_-

 \begin{equation}
     V_{\pm}:=\{x\in\mathbb{R}^4|x^2:=g_{\mu\nu}x^{\mu}x^{\nu}>0,\ \pm x^0>0\}.
 \end{equation}

%%%%%%%%%%%%%%%%%%%%%%%%%%%%%%%%%%%%%%%%%%%%
%%%%%%%%%%%%%%%%%%%%%%%%%%%%%%%%%%%%%%%%%%%%
%%%%%%%%%%%%%%%%%%%%%%%%%%%%%%%%%%%%%%%%%%%%
%%%%%%%%%%%%%%%%%%%%%%%%%%%%%%%%%%%%%%%%%%%%

 We will use the following short notation for products following \cite{bahns2003ultraviolet}. For $x_1,\cdots x_n\in\mathbb{R}^4$, $F\in\mathcal{E}(\mathbb{R}^{4n})$ and $G\in\mathcal{F}$

 \begin{equation}
     \begin{split}
         d\underline{x}^n&:=\prod_{j=1}^{n}\prod_{\mu=0}^{4}dx_{i}^{\mu},\\
         F(\underline{x})&:=F(x_1,\cdots x_n),\\
         \phi(\underline{x})^n&:=\phi(x_1)\cdots\phi(x_n),\\
         \frac{\delta^n G}{\delta \phi(\underline{x})^n}&:=\frac{\delta^n G}{\delta\phi(x_1)\cdots\delta\phi(x_n)}.
     \end{split}
 \end{equation}

 Fields with pointwise multiplication and involution given by complex conjugation, define the *-algebra of classical fields. Quantum fields are described by $\mathcal{F}\llbracket\hbar\rrbracket$, formal power series in the Planck constant with coefficients in $\mathcal{F}$; and with a non-commutative product. For $F,\ G\in\mathcal{F}$, the quantum product is

%%%%%%%%%%%%%%%%%%%%%%%%%%%%%%%%%%%%%%%%%%%%
%%%%%%%%%%%%%%%%%%%%%%%%%%%%%%%%%%%%%%%%%%%%
%%%%%%%%%%%%%%%%%%%%%%%%%%%%%%%%%%%%%%%%%%%%
%%%%%%%%%%%%%%%%%%%%%%%%%%%%%%%%%%%%%%%%%%%%
%Se cambio * por star

 \begin{equation}
    \label{starprod}
     F\star G=\sum_{n=0}^{+\infty}\frac{\hbar^n}{n!}\int d\underline{x}^nd\underline{y}^n\frac{\delta^n F}{\delta\phi(\underline{x})^n}\left(\Delta^+(\underline{x}-\underline{y})\right)^n\frac{\delta^n G}{\delta\phi(\underline{y})^n},
 \end{equation}

 %%%%%%%%%%%%%%%%%%%%%%%%%%%%%%%%%%%%%%%%%%%%
%%%%%%%%%%%%%%%%%%%%%%%%%%%%%%%%%%%%%%%%%%%%
%%%%%%%%%%%%%%%%%%%%%%%%%%%%%%%%%%%%%%%%%%%%
%%%%%%%%%%%%%%%%%%%%%%%%%%%%%%%%%%%%%%%%%%%%

 where 

 \begin{equation}
     \left(\Delta^+(\underline{x}-\underline{y})\right)^n:=\prod_{\ell=1}^{n}\Delta^+(x_{\ell}-y_{\ell}),
 \end{equation}

 with $\Delta^+$ the Wightman two-point function, which corresponds to the choice of positive frequency part of the Green function for the Klein Gordon operator $K=\Box+m^2$. The fields $\mathcal{F}\llbracket\hbar\rrbracket$ with this non-commutative product modulo the ideal $\mathbb{S}K_1\mathcal{F}\llbracket\hbar\rrbracket$; where $K_1$ \cite{chilian2009time} is the Klein-Gordon operator on the first variable and $\mathbb{S}$ is the symmetrization operator, define the \textit{on-shell} free scalar fields. Elements of $\mathcal{F}\llbracket\hbar\rrbracket$ are normally ordered fields and the quantum product is just the implementation of the Wick theorem.\\

 Another important non-commutative product in quantum field theory is the T-product (time ordered product). This product is fundamental on the definition of a perturbative Field Theory defined by an interaction Hamiltonian (e.g. causal S-matrix formalism of Bogoliubov-Epstein-Glaser \cite{scharf2014finite}). T-product defines elements which are formal power series in $\kappa$ (the coupling constant) and Laurent series in $\hbar$. From de Dyson series in perturbation theory for quantum mechanics, the S-matrix operator is

 \begin{equation}
     S=1+\sum_{n=1}^{+\infty}\frac{(-i\kappa)^n}{n!\hbar^n}\int dt_1\cdots dt_nT_n\left(H_{int}(t_1)\otimes\cdots\otimes H_{int}(t_n)\right),
 \end{equation}

%%%%%%%%%%%%%%%%%%%%%%%%%%%%%%%%%%%%%%%%%%%%
%%%%%%%%%%%%%%%%%%%%%%%%%%%%%%%%%%%%%%%%%%%%
%%%%%%%%%%%%%%%%%%%%%%%%%%%%%%%%%%%%%%%%%%%%
%%%%%%%%%%%%%%%%%%%%%%%%%%%%%%%%%%%%%%%%%%%%
%Se cambio * por star

 where $T_n\left(H_{int}(t_1)\otimes\cdots\otimes H_{int}(t_n)\right)=H_{int}(t_{\pi(1)})\star\cdots \star H_{int}(t_{\pi(n)})$ for $\pi\in S_n$ such that $t_{\pi(1)}>...>t_{\pi(n)}$. In QFT the T-product is exactly this time ordering for point $x_i\neq x_j$. For distributions with support in coinciding points, it is necessary to implement some renormalization procedure, i.e. extension of functionals from $\mathbb{R}^{4n}\backslash \Delta^{n}$ ($\Delta^{n}$ the thin diagonal) to the whole space.\\

 %%%%%%%%%%%%%%%%%%%%%%%%%%%%%%%%%%%%%%%%%%%%
%%%%%%%%%%%%%%%%%%%%%%%%%%%%%%%%%%%%%%%%%%%%
%%%%%%%%%%%%%%%%%%%%%%%%%%%%%%%%%%%%%%%%%%%%
%%%%%%%%%%%%%%%%%%%%%%%%%%%%%%%%%%%%%%%%%%%%

%%%%%%%%%%%%%%%%%%%%%%%%%%%%%%%%%%%%%%%%%%%%
%%%%%%%%%%%%%%%%%%%%%%%%%%%%%%%%%%%%%%%%%%%%
%%%%%%%%%%%%%%%%%%%%%%%%%%%%%%%%%%%%%%%%%%%%
%%%%%%%%%%%%%%%%%%%%%%%%%%%%%%%%%%%%%%%%%%%%
% Se cambio la o en Moller

 The relation between the T-product and the interacting theory is two-fold. In experiments in high energy physics, the role of the S-matrix is to connect asymptotic in and out states (which are assumed to be described by the free theory, far away from the support of the interaction before taking the adiabatic limit). On the other side, there are those retarded wave (M{\o}ller) operators $r_{S_1,S_0}$ \cite{dutsch2003master}, which map solutions of the free theory described by the free action

 %%%%%%%%%%%%%%%%%%%%%%%%%%%%%%%%%%%%%%%%%%%%
%%%%%%%%%%%%%%%%%%%%%%%%%%%%%%%%%%%%%%%%%%%%
%%%%%%%%%%%%%%%%%%%%%%%%%%%%%%%%%%%%%%%%%%%%
%%%%%%%%%%%%%%%%%%%%%%%%%%%%%%%%%%%%%%%%%%%%

 \begin{equation}
     S_0=\frac{1}{2}\int dx\left(\partial_{\nu}\phi\partial^{\nu}\phi-m^2\phi^2\right)
 \end{equation}

 to solutions of the interacting theory with action $S_1=S_0+\kappa S_{int}$

 \begin{equation}
     S_{int}[g]=-\int dx\ g(x)\mathcal{H}_{int}(x),
 \end{equation}

 where a function $g\in \mathcal{D}(\mathbb{R}^4)$ introduced in order to avoid IR divergences, and $\mathcal{H}_{int}$ is the interaction Hamiltonian density. The retarded operator can be defined as a perturbative series in $\kappa$ in terms of the R-product (retarded product). Let $F\in\mathcal{F}$

 \begin{equation}
     F^{ret}:=r_{S_1,S_0}(F)=\sum_{n=0}^{+\infty}\frac{1}{n!}R_{n,1}(S_{int}[g]\otimes...\otimes S_{int}[g],F).
 \end{equation}

 The R-product can be calculated in terms of the T-product using the Bogoliubov formula

%%%%%%%%%%%%%%%%%%%%%%%%%%%%%%%%%%%%%%%%%%%%
%%%%%%%%%%%%%%%%%%%%%%%%%%%%%%%%%%%%%%%%%%%%
%%%%%%%%%%%%%%%%%%%%%%%%%%%%%%%%%%%%%%%%%%%%
%%%%%%%%%%%%%%%%%%%%%%%%%%%%%%%%%%%%%%%%%%%%
%Se cambio * por star

 \begin{align}
     F^{ret}=R\left(e^{\kappa S_{int}/\hbar}_{\otimes},F\right)&=\frac{\hbar}{i}\frac{d}{d\lambda}\Big|_{\lambda=0}\textbf{S}(\kappa S_{int}/\hbar)^{\star -1}\star\textbf{S}(\kappa S_{int}/\hbar+\lambda F),\\
     R_n(S\otimes...\otimes S,F)&=i^n\sum_{I\subseteq\{1,...,n\}}(-1)^{|I|}\bar{T}_{|I|+1}(S_I\otimes F)\star T_{|I^c|}(S_{I^c})
 \end{align}

 %%%%%%%%%%%%%%%%%%%%%%%%%%%%%%%%%%%%%%%%%%%%
%%%%%%%%%%%%%%%%%%%%%%%%%%%%%%%%%%%%%%%%%%%%
%%%%%%%%%%%%%%%%%%%%%%%%%%%%%%%%%%%%%%%%%%%%
%%%%%%%%%%%%%%%%%%%%%%%%%%%%%%%%%%%%%%%%%%%%

 with S matrix

 \begin{equation}
     \textbf{S}(F)=1+\sum_{n=1}^{+\infty}\frac{i^n}{n! \hbar^n}T_n(F^{\otimes n})=T\left(e_{\otimes}^{iF/\hbar}\right),
 \end{equation}

 where the exponential with respect to the tensor product is defined by

 \begin{equation}
     e_{\otimes}^F:=1+\sum_{n=1}^{+\infty}\frac{1}{n!}\underbrace{F\otimes...\otimes F}_{n-times}=1+\sum_{n=1}^{+\infty}\frac{1}{n!}F^{\otimes n},
 \end{equation}

 and anti chronological ($\bar{T}$-) product given by the inverse order of the T-product.

\section{Quantum Fields in Quantum Spacetime}
\label{sec:4}

 The aim of this section is to associate to certain class of fields in standard pAQFT, new fields depending on the non-commutativity parameter $\lambda_P$, using the properties discussed for the algebra of functions on the QST in the definition. This new fields should be applied in perturbation theory (e.g. Dyson series, retarded fields) for the associated interacting effective Hamiltonians for interactions of the form $\phi^n$ \cite{peskin2018introduction}.\\

 Given that, from the second section, the available product over the QST is defined for functions with the same coordinate generator $q$, the aforementioned association in the last paragraph is expected to be a map from \textit{local} fields $\mathcal{F}_{loc}$ to fields $\mathcal{F}$, where the image shall be understood as the effective field over the QST. As it is customary in approaches to QFT from non-commutative geometry, e.g. the Moyal plane \cite{akofor2008quantum}; if the departure point of the interaction is a local action, the effective action taking into account the non-commutativity effects results non-local\cite{bahns2002unitarity,bahns2004perturbative,namsrai1986nonlocal}.\\

 For this reason, during the following sections, the fields of interest are assumed to be local, i.e. the support of the distributions which define the fields shall be in the thin diagonal

 \begin{equation}
     supp\ f_n\subseteq \Delta^n.
 \end{equation}

 A quite general set of examples of those fields are of the form

 \begin{equation}
     \begin{split}
         F[\phi]&=\sum_{n=1}^{N}\sum_{\substack{a=(a_1,...,a_n)\in\mathbb{N}^d\\
        |a|<\infty}}\int d\underline{x}^n(-1)^{|a|}\partial^a\left[f_{a}(x_n)\delta(x_1-x_n)...\delta(x_{n-1}-x_n)\right]\phi(\underline{x})^n\\
        &=\sum_{n=1}^{N}\sum_{\substack{a=(a_1,...,a_n)\in\mathbb{N}^d\\
        |a|<\infty}}\int dx\ f_{a}(x)\underline{\partial^{a}\phi(x)},
     \end{split}
 \end{equation}

 where

 \begin{equation}
     \begin{split}
         |a|&:=|a_1+...+a_n|,\\
         \partial^a&:=\partial_{x_1}^{a_1}...\partial_{x_n}^{a_n},\\
         \underline{\partial^{a}\phi(x)}&:=\partial^{a_1}\phi(x)\cdots\partial^{a_n}\phi(x),
     \end{split}
 \end{equation}

 and $f_{a_1...a_n}$ are compactly supported distributions in $\mathbb{R}^4$. An example of this kind of local fields is the $\phi^n$ interaction

 \begin{equation}
     S_{int}[g]=-\frac{1}{n!}\int dx\ g(x)\phi(x)^n,
 \end{equation}

 where $g\in\mathcal{D}(\mathbb{R}^4)$. With the definition of the derivative of a function over the QST

 \begin{equation}
     \frac{\partial}{\partial x^{\mu}}f(q)=\frac{\partial}{\partial a^{\mu}}\Big|_{a^{\mu}=0}f(q+a^{\mu}I)=\int\frac{dk}{(2\pi)^{4}}\hat{f}(Q,k)(ik_{\mu})e^{ik_{\mu}q^{\mu}}.
 \end{equation}

 The respective fundamental quantum field over the quantum spacetime is defined by

 \begin{equation}
     \phi(q)=\int\frac{dk}{(2\pi)}\hat{\phi}(k)\otimes e^{ik_{\mu}q^{\mu}},
 \end{equation}

 where the Fourier transform of the field is the usual

 \begin{equation}
     \hat{\phi}(k)=\int dx\ \phi(x)e^{-ik_{\mu}x^{\mu}}.
 \end{equation}

 This is an element of the tensor product of *-algebras $\mathcal{F}\otimes\mathcal{E}$. In the following, the $\otimes$ symbol shall be omitted. Making use of the product in $\mathcal{E}$, the monomial field over the QST is

 \begin{equation}
     \underline{\partial^{a}\phi(q)}=\int\frac{d\underline{k}^n}{(2\pi)^{4n}}exp\left\{i\frac{\lambda_P^2}{2}\sum_{j<m}k^{j}_{\mu}Q^{\mu\nu}k^{m}_{\nu}\right\}(i\underline{k})^{a}\hat{\phi}(\underline{k})^n e^{i\sum_{j}k^{j}_{\mu}q^{\mu}},
 \end{equation}

 where $(ik)^a=\prod_{\mu=0}^{3}(ik_{\mu})^{a^{\mu}}$ for $a=(a^{0},...,a^{3})\in\mathbb{N}^{4}$. Elements $\Phi\in\mathcal{F}\otimes\mathcal{E}$ can be seen as maps from $\mathcal{E}'$ (functionals over the QST, such as the states $\mathcal{S}(\mathcal{E}))$ to fields in $\mathcal{F}$. For a general element of the form

 \begin{equation}
     \Phi(q)=\int\frac{dk}{(2\pi)^4}\hat{\Phi}(Q,k)e^{ik_{\mu}q^{\mu}},
 \end{equation}

 and a functional $\omega=\int_{\Sigma}d\mu(\sigma)\omega_{\sigma}\in\mathcal{E}'$, a field is defined by the pairing

 \begin{equation}
     \langle\Phi,\omega\rangle=\int_{\Sigma}d\mu(\sigma)\omega_{\sigma}\left(e^{ik_{\mu}\pi_{\sigma}(q^{\mu})}\right)\hat{\Phi}(\sigma,k).
 \end{equation}

 In particular, it is possible to define a generalization of the integral from the calculus at the end of section II, to define the action of a compactly supported distribution $f\in\mathcal{D}'(\mathbb{R}^4)$. Define the linear functional $I(f)\in\mathcal{E}'$ with the rotationally invariant measure $\mu$ on $\Sigma_1$ and its action on irreducible representations

 \begin{equation}
     I(f)_{\sigma}\left(e^{ik_{\nu}\pi_{\sigma}(q^{\nu})}\right)=\hat{f}(-k).
 \end{equation}

 Then, given the local field $F[\phi]\in\mathcal{F}_{loc}$; the effective non-local field $F^{(Q)}[\phi]\in\mathcal{F}$ is defined in the momentum space as

 \begin{equation}
        F^{(Q)}[\phi]=\sum_{n=1}^{N}\sum_{\substack{a=(a_1,...,a_n)\in\mathbb{N}^d\\
        |a|<\infty}}\langle\underline{\partial^{a}\phi(q)},I(f_{a})\rangle,
 \end{equation}

 where

 \begin{equation}
     \begin{split}
     \langle\underline{\partial^{a}\phi(q)},I(f)\rangle&=\int_{\Sigma_1}d\mu(\sigma)\int\frac{d\underline{k}^n}{(2\pi)^{4n}}\hat{f}\left(-\sum_{j=1}^{n}k^j\right)(i\underline{k})^{a}exp\left\{i\frac{\lambda_P^2}{2}\sum_{j<l}k^j_{\mu}\sigma^{\mu\nu}k^l_{\nu}\right\}\hat{\phi}(\underline{k})^n
     \end{split}
 \end{equation}

 For example, the effective action for the $\phi^n$ interaction action obtained by the previous map, in the momentum space, is

 \begin{equation}
     S^{(Q)}_{int}[g]=-\frac{1}{n!}\int_{\Sigma_1}d\mu(\sigma)\int\frac{d\underline{k}^n}{(2\pi)^{4n}}\hat{g}\left(-\sum_j k^j\right)exp\left\{i\frac{\lambda_P^2}{2}\sum_{j<m}k^{j}_{\mu}\sigma^{\mu\nu}k^{m}_{\nu}\right\}\hat{\phi}(\underline{k})^n
 \end{equation}

 In order to obtain explicit formulas for the action, as well as  an interacting Hamiltonian, let us  first calculate the integration term associated with the non-commutativity parameter \cite{doplicher1995quantum}. That term is defined by

 \begin{equation}
    \label{eqlambda1}
     \begin{split}
         \Lambda_n(\underline{k})&=\int_{\Sigma_1}d\mu(\sigma)exp\left\{i\frac{\lambda_P^2}{2}\sum_{j<m}k^{j}_{\mu}\sigma^{\mu\nu}k^{m}_{\nu}\right\}\\
         &=\frac{1}{2}\left(\frac{sin(\beta_+)}{\beta_+}+\frac{sin(\beta_-)}{\beta_-}\right),
     \end{split}
 \end{equation}

 where

 \begin{equation}
    \label{eqlambda2}
     \beta_{\pm}:=\frac{\lambda_P^2}{2}\left\lVert\sum_{j<l}k_0^j\vec{k}^l-k_0^l\vec{k}^j\pm\vec{k}^j\times\vec{k}^l\right\rVert.
 \end{equation}

 Its inverse Fourier transform is denoted by

 \begin{equation}
     \tilde{\Gamma}_n(\underline{x})=\int\frac{d\underline{k}^n}{(2\pi)^{4n}}\Lambda_n(\underline{k})e^{i\sum_j k^j_{\mu}x_j^{\mu}};
 \end{equation}

 and the translation to the point $x\in\mathbb{R}^4$

 \begin{equation}
     \Gamma_n(\underline{x};x)=\tilde{\Gamma}_n(\underline{x}-x),
 \end{equation}

 where

 \begin{equation}
     \tilde{\Gamma}_n(\underline{x}-x)=\tilde{\Gamma}_n(x_1-x,\cdots x_n-x).
 \end{equation}

 Completes the definition of the non-commutative action in terms of a non-local interaction distribution $\Gamma$, as follows

 \begin{equation}
    \label{inter}
     S^{(Q)}_{int}[g]=-\frac{1}{n!}\int dx\ d\underline{x}^n g(x)\Gamma(\underline{x};x)\phi(\underline{x})^n.
 \end{equation}

 For the general local field $F[\phi]$, the action of this map defines a non-local effective field, which in the configuration space if given by

 \begin{equation}
     \langle\underline{\partial^{a}\phi(q)},I(f)\rangle=\int dxd\underline{x}^n\Gamma_n(\underline{x};x)\underline{\partial^{a}\phi(x)}
 \end{equation}

 It is straightforward that in the weak limit when $\lambda_P\rightarrow 0$, the non-local kernel

 \begin{equation}
     \lim_{\lambda_P\rightarrow 0}\Gamma(x_1,\ldots,x_n;x)=\delta(x_1-x)\cdots\delta(x_n-x),
 \end{equation}

 and the standard QFT is recovered. The effective action above $S^{(Q)}_{int}$ can be understood as the associated action of some perturbative QFT induced by the non-local Hamiltonian density

 \begin{equation}
     \mathcal{H}_{int}(x)=\frac{1}{n!}\int d\underline{x}^n\Gamma_n(\underline{x};x)\phi(\underline{x})^n.
 \end{equation}

 As in the previous section, the \textbf{S} matrix is defined in terms of the T-products, linear maps

 \begin{equation}
     T_n:\mathcal{F}^{\otimes n}\rightarrow \mathcal{F}\llbracket\kappa,\hbar,\hbar^{-1}\rrbracket,
 \end{equation}

 which, roughly speaking, order the fields $\phi$ with respect to the time coordinate; but as in the definition from Section III for standard QFT, the time ordering is performed with respect to the time coordinate $x$ of the Hamiltonian density, in this case ignoring the \textit{dummy coordinates} integrated with the non-local kernel $\Gamma$. For this reason, the $x$ coordinate is referred as \textit{time stamp} . For example, at order $N=2$ in the interaction coupling, the S matrix is

 \begin{align*}
     S[g]=&1-\frac{i\kappa}{n!\hbar}\int\ dxd\underline{x}g(x)\Gamma_n(\underline{x};x)\phi(\underline{x})^n\\
     &-\frac{\kappa^2}{2(n!)^2\hbar^2}\int dx_{1}dx_{2}d\underline{x_1}^nd\underline{x_2}^ng(x_1)g(x_2)T_2\big(\Gamma_n(\underline{x_1};x_1)\phi(\underline{x_1})^n\\
     &\otimes\Gamma_n(\underline{x_2};x_2)\phi(\underline{x_2})^n\big)+\mathcal{O}(\kappa^3),
 \end{align*}

 for $x_1^0\neq x_2^0$, $\underline{x_i}=(x_{i,1},\cdots,x_{i,n})$ for $i=1,2$ and

%%%%%%%%%%%%%%%%%%%%%%%%%%%%%%%%%%%%%%%%%%%%
%%%%%%%%%%%%%%%%%%%%%%%%%%%%%%%%%%%%%%%%%%%%
%%%%%%%%%%%%%%%%%%%%%%%%%%%%%%%%%%%%%%%%%%%%
%%%%%%%%%%%%%%%%%%%%%%%%%%%%%%%%%%%%%%%%%%%%
%Se cambio * por star

 \begin{align*} T_2\big(\Gamma_n&(\underline{x_1};x_1)\phi(\underline{x_1})^n\otimes  \Gamma_n(\underline{x_2};x_2)\phi(\underline{x_2})^n\big)\\
 =&\Gamma_n(\underline{x_1};x_1)
    \Gamma_n(\underline{x_2};x_2)\big(\theta(x_1^0-x_2^0)\phi(\underline{x_1})^n\star\phi(\underline{x_2})^n\\
    &+\theta(x_2^0-x_1^0)\phi(\underline{x_2})^n\star\phi(\underline{x_1})^n\big).
 \end{align*}

 %%%%%%%%%%%%%%%%%%%%%%%%%%%%%%%%%%%%%%%%%%%%
%%%%%%%%%%%%%%%%%%%%%%%%%%%%%%%%%%%%%%%%%%%%
%%%%%%%%%%%%%%%%%%%%%%%%%%%%%%%%%%%%%%%%%%%%
%%%%%%%%%%%%%%%%%%%%%%%%%%%%%%%%%%%%%%%%%%%%

 For both T- and R- products, it is possible to obtain a set of Feynman rules modified by the non-local effects of the effective interacting QFT, the latter one introducing the T-product into the Bogoliubov formula.\\

 As an example of the previous formulas, let us calculate the interacting field $\phi_S(x)$, defined as a formal power series in $\kappa$ and $\hbar$ for $n=4$ order by order as

\begin{equation}
    \begin{split}
        \phi_S(x):=R\left(e_{\otimes}^{\kappa S_I[g]/\hbar},\phi(x)\right)=\phi(x)+\frac{\kappa}{\hbar}R_{1,1}\left(S_I[g],\phi(x)\right)\\
        +\frac{\kappa^2}{2\hbar^2}R_{2,1}\left(S_I[g]\otimes S_I[g],\phi(x)\right)+\mathcal{O}(\kappa^3),
    \end{split}
\end{equation}

keeping concern to the 2nd order approximation in $\kappa$. The zeroth order is simply $\phi_S^{(0)}(x)=\phi(x)$, the free field. The first order term is

\begin{equation}
    \frac{\kappa}{\hbar}R_{1,1}\left(S_I[g],\phi(y)\right)=\frac{\kappa}{4!}\int dxd\underline{x}^4g(x)\Gamma_4(\underline{x};x)\theta(y^0-x^0)\sum_{j=1}^{4}\Delta_m(y-x_j)\prod_{\substack{k=1\\k\neq j}}^{4}\phi(x_k),
\end{equation}

using from the Bogoliubov formula that

\begin{equation}
    R_{1,1}(S,F)=i\left[T_2(S,F)-S\star_m F\right]
\end{equation}

As it was mentioned above, the time order (in the $\theta$ functions) is related only between the time stamps $y^0$ and  $x^0$, but not the dummy variables $x_1,\ldots,x_n$; and $\Delta_m(z)=\Delta_m^+(z)-\Delta_m^+(-z)$ is the commutator (Pauli-Jordan) function. The latter term can be represented as the Feynman diagram at Fig. 1.\\

\begin{figure}
\centerline{\includegraphics[width=4cm]{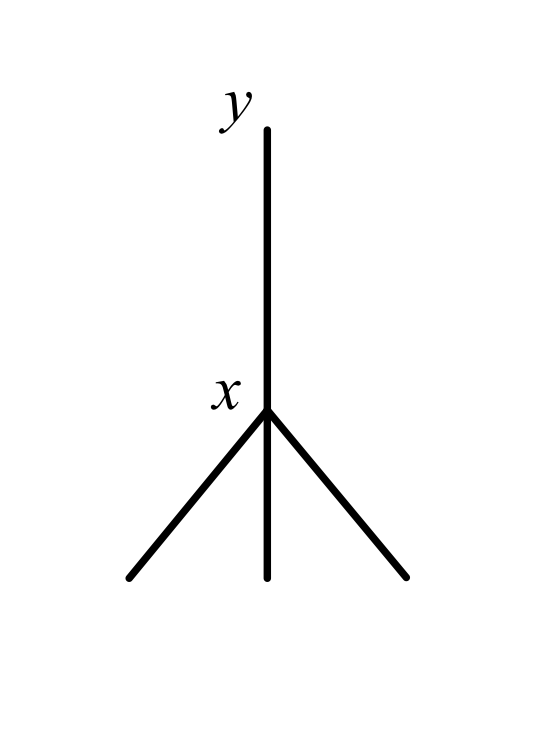}}
\vspace*{-20pt}
\caption{A one-vertex tree level Feynman diagram for a $\phi^4$-interaction.}
\end{figure}

Assuming that $\Gamma_4$ is symmetrized in $x_1,\ldots,x_4$; it corresponds to a normally orderded quantum field of the form

\begin{equation}
    \int dx\,dx_1\cdots dx_{3}f^{(3)}(x_1,\ldots,x_{3};y)\phi(x_1)\cdots\phi(x_{3})
\end{equation}

with the distribution in $\mathcal{D}'(\mathbb{R}^{4\times 3})$

\begin{equation}
    f^{(3)}(x_1,...,x_{3};y)=\kappa \int dxdx_4 g(x)\Gamma_n(\underline{x};x)\theta(y^0-x^0)\Delta_m(y-x_4)
\end{equation}

This corresponds to a tree diagram, then this term is due to classical interacting field contribution over non-commutative spacetime. The second order contribution results more involved because it requires to calculate

\begin{equation}
    \begin{split}
        R_{2,1}(S_1\otimes S_2,F)=T_3(S_1\otimes S_2\otimes F)-S_1\star T_2(S_2\otimes F)+\\
        -S_2\star T_2(S_1\otimes F)-T_2(S_1\otimes S_2)\star F+S_1\star S_2\star F+S_2\star S_1\star F
    \end{split}
\end{equation}

For higher order contributions, the calculation of several terms leads to the following Feynman rules for the interacting field $\phi_S(x)$ at order $k$ in $\kappa$ are as follows:

\begin{itemize}
    \item Draw every diagram with one vertex $x$ (with only one leg) and $k$ vertices labeled $x_1,\ldots, x_k$ which are the time stamps. Also for each of the latter labels there will be required $x_{i.j}$ non-local variables (which are not time-ordered) for $i\in\{1,\ldots,k\}$ and $j\in\{1,\ldots,n\}$. Also, pick up a temporal order $x^0_{\pi(1)}<\cdots <x^0_{\pi(k)}$ for $\pi\in S_k$.
    \item For the line connecting $x$ with $x_i$, put a factor
    \begin{equation}
        \theta(x^0-x^0_i)\sum_{j=1}^{4}\Delta_m(x-x_{i.j}).
    \end{equation}
    Inside the sum $x_{i.j}$ is an occupied label.
    \item For every $t$ lines connecting $x_i$ with $x_j$ with $x^0_i>x^0_j$, put a factor inside the previous sums put a factor
    \begin{equation}
        \theta(x^0_i-x^0_j)\sum_{m_1,\ldots,m_t}\sum_{n_1,\ldots,n_t}\left(\prod_{\ell=1}^t\Delta_m^+\left(x_{i.m_{\ell}}-x_{j.n_{\ell}}\right)-\prod_{\ell=1}^t\Delta_m^+\left(x_{j.n_{\ell}}-x_{i.m_{\ell}}\right)\right),
    \end{equation}
    where $x_{i.m_{\ell}}$ and $x_{j.n_{\ell}}$ are taken over every label which is not occupied. The labels inside $\Delta_m^+$ in the sum are now also occupied.
    \item Once the steps above are completed, for each free leg in vertex $x_i$, a field $\phi(x_{i,j})$ is multiplied, with $j$ taking values over each label which is not occupied.
    \item If $L$ is the number of lines, the last expression is, up to some symmetry numerical factor, the integrand of
    \begin{equation}
        \kappa^k\hbar^L\int d\underline{x}^kd\underline{x_1}^n\cdots d\underline{x_k}^n g(\underline{x})^k\Gamma_n(\underline{x_1};x)\cdots\Gamma_n(\underline{x_k};x_k)
    \end{equation}
    \item Sum over all orders $\pi\in S_k$ and over all diagrams. This is $R_{k,1}$
\end{itemize}

\textit{Example: }Let us consider the contributions of all the diagrams of the form represented in Fig. 2 for $n=4$.

\begin{figure}
\centerline{\includegraphics[width=5cm]{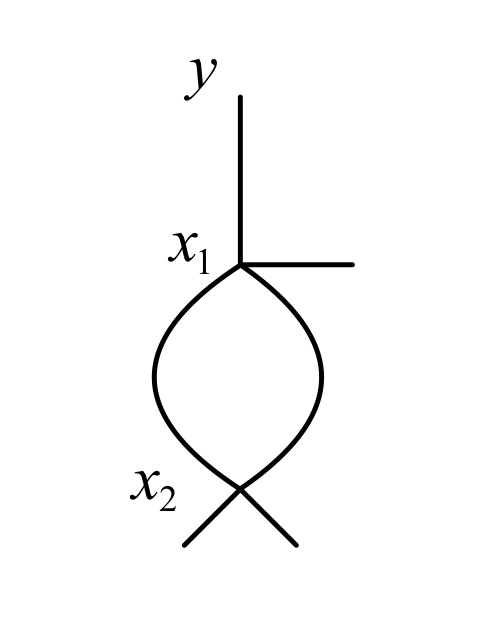}}
\vspace*{-8pt}
\caption{A two-vertex one loop Feynman diagram for a $\phi^4$-interaction.}
\end{figure}

This is a $k=2$ and $L=3$ contribution to the interacting field. The field associated with this diagram for $x_1^0>x_2^0$ is

\begin{align*}
    &\frac{\kappa^2\hbar}{4(4!)^2}\int d\underline{x}^2d\underline{x_1}^4d\underline{x_2}^4g(\underline{x})^2\Gamma_4(\underline{x_1};x_1)\Gamma_4(\underline{x_2};x_2)\times\\
    &\times\Bigg\{\theta(x^0-x_1^0)\theta(x_1^0-x_2^0)\sum_{j=1}^4\Delta_m(y-x_{1.j})\sum_{\substack{j_1=1\\j_1\neq j}}^4\sum_{\substack{j_2=1\\j_2\neq j,j_1}}^4\sum_{k_1=1}^4\sum_{\substack{k_2=1\\k_1\neq k_1}}^4\Bigg(\prod_{t=1}^2\Delta_m^+(x_{1.j_t}-x_{2.k_t})\\
    &-\prod_{t=1}^2\Delta_m^+(x_{2.k_t}-x_{1.j_t})\Bigg)\prod_{\substack{m=1\\m\neq j,j_1,j_2}}^4\phi(x_{1.m})\prod_{\substack{n=1\\n\neq k_1,k_2}}^4\phi(x_{2.n})\Bigg\}.
\end{align*}

In any case, the fields and the propagators are mixed, due to the non-local kernels $\Gamma_4$, at different points of the spacetime.

\section{Unitarity of the S matrix}
\label{sec:5}

Usual properties of the $S$ matrix are not guaranteed when basic assumptions of standard QFT are lifted. For example, it is expected that the non-local effective action spoils causality for the interacting n-point functions; but causality is replaced by by the causality order induced by the time stamps, which comes from the choice of the state of the spacetime part of the algebra which manifestly breaks Lorentz invariance. Nonetheless, causality violation occurs at most at first order in $\lambda_P$, as taking its limit to zero coincides with standard local $\phi^n$ interaction, as this is an effect due to non commutativity of the spacetime.\\

Something different happens with unitarity of the S matrix. There are known cases of unitarity violation in QFT implementations on non commutative spacetimes, like the Moyal plane \cite{alvarez2001remarks,gomis2000space}. The lack of unitarity in those cases comes from keeping the non commutativity of spacetime coordinate operators at the moment of calculating the n-point functions. This leads to failure at some orders in $\kappa$ and $\hbar$ in perturbation theory of the unitarity condition:

\begin{equation}
    S[g]^{*}\star S[g]=1=S[g]\star S[g]^{*}.
\end{equation}

For standard QFT at second order in both $\kappa$ and $\hbar$, unitarity condition as it is discussed in \cite{bahns2002unitarity} takes the form

\begin{equation}
    (\Delta^F)^2+(\overline{\Delta^F})^2=(\Delta^+)^2+(\Delta^-)^2.
\end{equation}

Our choice of the time order induced by the state of the spacetime is compatible with the $\star$ product of quantum fields. This can be seen explicitly because for real interactions, involution changes time-ordered T-product by antichronological order $\overline{T}$. As evaluation $\phi(x)$ is taken on real functions, and $\Gamma_n$ is a real distribution (maps real valued function into real valued functions); in the definition of the S-matrix is equivalent to take conjugation and change chronological by antichronological products or changing $\Delta^+$ by $\Delta^-$ for the $\star$-product. Using

\begin{equation}
    \label{unitsec}
    (\Delta^+)(z)^{*}=\Delta^+(-z)=:\Delta^-(z),
\end{equation}

the $\overline{\star}$-product is defined by exchanging $\Delta^+$ by $\Delta^{-}$ in the definition of the product \eqref{starprod}. In order to observe in more detail that unitarity of the S matrix is given by similar arguments as in standard QFT, let us take a look up to second order in the interaction $\kappa$. At zeroth order it is trivial, and at first order it is just the fact that the effective interaction is real (Hermitian)

\begin{equation}
    \begin{split}
        &\left(\frac{i\kappa}{\hbar n!}\right)\int dx\ d\underline{x}\ g(x)\Gamma_{n}(\underline{x};x)\phi^{\otimes n}(\underline{x})+\\
        &+\left(\frac{-i\kappa}{\hbar n!}\right)\int dx\ d\underline{x}\ g(x)\Gamma_{n}(\underline{x};x)\phi^{\otimes n}(\underline{x})=0.
    \end{split}
\end{equation}

At second order in $\kappa$

\begin{equation}
    \begin{split}
        &\left(\frac{i\kappa}{\hbar n!}\right)^2\int dx_1 dx_2 d\underline{x_1} d\underline{x_2} g(x_1) g(x_2)\Gamma_{n}(\underline{x_1};x_1)\Gamma_n(\underline{x_2};x_2)\theta(x_1^0-x_2^0)\phi^{\otimes n}(\underline{x_1})\star\phi^{\otimes n}(\underline{x_2})+\\
        &+\left(-\frac{i\kappa}{\hbar n!}\right)^2\int dx_1 dx_2 d\underline{x_1} d\underline{x_2} g(x_1) g(x_2)\Gamma_{n}(\underline{x_1};x_1)\Gamma_n(\underline{x_2};x_2)\theta(x_1^0-x_2^0)\phi^{\otimes n}(\underline{x_1})\overline{\star}\phi^{\otimes n}
        (\underline{x_2})+\\
        &-\left(\frac{i\kappa}{\hbar n!}\right)^2\int dx_1 dx_2 d\underline{x_1} d\underline{x_2} g(x_1) g(x_2)\Gamma_{n}(\underline{x_1};x_1)\Gamma_n(\underline{x_2};x_2)\phi^{\otimes n}(\underline{x_1})\star\phi^{\otimes n}(\underline{x_2}),
    \end{split}
\end{equation}

where the last term is the $\star$- product of first order terms. The last expression is clearly zero by interchanging $(\underline{x_1};x_1)$ and $(\underline{x_2};x_2)$ in the middle term. This takes back $\Delta^-$ to $\Delta^+$ and the $\overline{\star}$ goes to $\star$, but changing $\theta(x_1^0-x_2^2)$ to $\theta(x_2^0-x_1^2)=1-\theta(x_1^0-x_2^2)$. This is a result that happens at every order in perturbation theory due to formal unitarity of the S-matrix. As it is done in \cite{bahns2002unitarity}, at second order in $\hbar$

\begin{equation}
    \begin{split}
        &\theta(x_1^0-x_2^0)\sum_{\{i_1,i_2\},\{j_1,j_2\}\subset \{1,...,n\}}\Delta^{+}(x_{1.i_1}-x_{2.j_1})\Delta^{+}(x_{1.i_2}-x_{2.j_2})+\\
        &+\theta(x_2^0-x_1^0)\sum_{\{i_1,i_2\},\{j_1,j_2\}\subset \{1,...,n\}}\Delta^{-}(x_{1.i_1}-x_{2.j_1})\Delta^{-}(x_{1.i_2}-x_{2.j_2})+\\
        &\theta(x_1^0-x_2^0)\sum_{\{i_1,i_2\},\{j_1,j_2\}\subset \{1,...,n\}}\Delta^{-}(x_{1.i_1}-x_{2.j_1})\Delta^{-}(x_{1.i_2}-x_{2.j_2})+\\
        &+\theta(x_2^0-x_1^0)\sum_{\{i_1,i_2\},\{j_1,j_2\}\subset \{1,...,n\}}\Delta^{+}(x_{1.i_1}-x_{2.j_1})\Delta^{+}(x_{1.i_2}-x_{2.j_2})=\\
        &=\sum_{\{i_1,i_2\},\{j_1,j_2\}\subset \{1,...,n\}}\Delta^{+}(x_{1.i_1}-x_{2.j_1})\Delta^{+}(x_{1.i_2}-x_{2.j_2}+\\
        &\sum_{\{i_1,i_2\},\{j_1,j_2\}\subset \{1,...,n\}}\Delta^{-}(x_{1.i_1}-x_{2.j_1})\Delta^{-}(x_{1.i_2}-x_{2.j_2});
    \end{split}
\end{equation}

is a non-commutative version of the second order case of unitarity \eqref{unitsec}. But this is a result of $\theta(x_1^0-x_2^0)+\theta(x_2^0-x_1^0)=1$ just as in second order for the unitarity of the S matrix in standard QFT.

\section{Singularity structure of the non-local kernel $\Gamma_n$}
 \label{sec:6}

In standard perturbation theory for QFT, when the expansion in terms of $\kappa$ and $\hbar$ is performed, divergences in Fourier transform space appear commonly, and are addressed with regularization techniques like analytic regularization \cite{velo1975renormalization} (e.g. dimensional regularization \cite{van2003introduction}) or momentum space cut-off \cite{peskin2018introduction}. With closer attention to the obtained distributions in Minkowski (physical) space-time from the Feynman diagrams, the resulting divergences can be traced back to ill-defined products of distributions. Bogoliubov-Epstein-Glaser renormalization \cite{epstein1973role,scharf2014finite} deals with the problem of defining those products in two steps. First, for functions supported where the product of distributions is well defined, i.e. where H\"ormander criterion is satisfied \cite{hormander2015analysis} (as it will be explained below), the usual product is taken. Then, the restricted distribution is extended (in a non necessarily unique way) to a distribution to the whole spacetime, with suitable scaling properties, known as the scaling and mass expansion axiom \cite{dutsch2019classical}. In this section, the wavefront set for the distributions involved in our modified Feynman rules will be presented. This set will allow us to describe on which points the products of distributions are well defined.

There are simple cases when the product of distributions can be defined \cite{hormander2015analysis}. Given a distribution $\Gamma\in\mathcal{D}'(\mathbb{R}^n)$; we say that $\Gamma$ is a smooth function restricted to the open set $X\subseteq\mathbb{R}^n$ if there is a smooth function $\psi_{\Gamma}$ such that for any $\phi\in\mathcal{D}(X)$

\begin{equation}
    \Gamma(\phi)=\int dx\psi_{\Gamma}(x)\phi(x).
\end{equation}

The complement of the biggest $X$ where this is possible, the set where $\Gamma$ fails to be a function, is called the singular support of $\Gamma$ and is denoted by $sing\ supp(\Gamma)$. It is clear when the singular supports of $\Gamma_1$ and $\Gamma_2$ are disjoint, the product is well defined: given a smooth function $\chi$ which is 1 at $sing\ supp(\Gamma_1)$ and 0 at $sing\ supp(\Gamma_2)$, for any $\phi\in\mathcal{D}$

\begin{equation}
    (\Gamma_1\cdot\Gamma_2)(\phi)=\Gamma_1(\psi_{\Gamma_2}\chi\psi)+\Gamma_2(\psi_{\Gamma_1}(1-\chi)\phi).
\end{equation}

The Fourier transform of a compactly supported distribution $\Gamma\in\mathcal{E}'(\mathbb{R}^n)$ is

\begin{equation}
    \hat{\Gamma}(k)=\Gamma(e^{-ik_{\mu}x^{\mu}}).
\end{equation}

From the Paley-Wiener theorem \cite{hormander2015analysis}, another characterization when a distribution is a function is given in terms of the decay properties of its Fourier transform. $\Gamma$ is a distribution if and only if $\hat{\Gamma}$ decays faster than any inverse power of $k$. The set that measure the failure of a distribution to be a smooth function in the Fourier space, in analogy to the singular support, is the \textit{cone}.\\

%%%%%%%%%%%%%%%%%%%%%%%%%%%%%%%%%%%%%%%%%%%%
%%%%%%%%%%%%%%%%%%%%%%%%%%%%%%%%%%%%%%%%%%%%
%%%%%%%%%%%%%%%%%%%%%%%%%%%%%%%%%%%%%%%%%%%%
%%%%%%%%%%%%%%%%%%%%%%%%%%%%%%%%%%%%%%%%%%%%
% Se aclara que los vectores en el cono Sigmo no pueden incluir el cero y se adiciona un comentario que, aunque la transformada de Fourier esta definida para distribuciones con soporte compacto, el WF set està definido para distribuciones 

The cone of a distribution, denoted as $\Sigma(\Gamma)$, is the complement of largest open conical neighborhood in $\mathbb{R}^n\backslash \{0\}$ where $\hat{\Gamma}$ decays faster than any inverse polynomial function. With these definitions, the main idea of microlocal analysis is to define locally, the information of the singular behavior described by $\Sigma(\Gamma)$. For $x\in\mathbb{R}^n$, the cone at $x$ of $\Gamma\in\mathcal{D}'(\mathbb{R}^n)$ is

\begin{equation}
    \Sigma_{x}(\Gamma):=\bigcap_{\substack{\phi\in\mathcal{D}(\mathbb{R}^n)\\x\in supp(\phi)}}\Sigma(\phi\Gamma).
\end{equation}

Observe that although $\Gamma\in\mathcal{D}'(\mathbb{R}^n)$, for $\phi\in\mathcal{D}$, the product $\phi\Gamma$ defined by $(\phi\Gamma)(\psi)=\Gamma(\phi\cdot\psi)$ for any $\psi\in\mathcal{E}(\mathbb{R}^n)$ is a compactly supported distribution, then its Fourier transform and cone at any point are well defined. The wavefront set $WF(\Gamma)$ encodes the information of both $sing\  supp$ and $\Sigma$. It is defined by

%%%%%%%%%%%%%%%%%%%%%%%%%%%%%%%%%%%%%%%%%%%%
%%%%%%%%%%%%%%%%%%%%%%%%%%%%%%%%%%%%%%%%%%%%
%%%%%%%%%%%%%%%%%%%%%%%%%%%%%%%%%%%%%%%%%%%%
%%%%%%%%%%%%%%%%%%%%%%%%%%%%%%%%%%%%%%%%%%%%

\begin{equation}
    WF(\Gamma)=\left\{(x,k)\in\mathbb{R}^n\times\mathbb{R}^n|x\in sing\ supp(\Gamma),\ k\in\Sigma_x(\Gamma)\right\},
\end{equation}

and for distributions over a manifold $M$, it can be defined as a subbundle of $T^*M$. The H\"ormander's criterion establishes that the product of two distributions $\Gamma_1\cdot\Gamma_2$ is a well defined distribution if the pointwise sum $WF(\Gamma_1)+WF(\Gamma_2)$ does not intersect the zero section, i.e.

\begin{equation}
    \left\{(x,k_1+k_2)|(x,k_1)\in WF(\Gamma_1)\ and\ (x,k_2)\in WF(\Gamma_2)\right\}\cap \mathbb{R}^n\times\{0\}=\emptyset.
\end{equation}

%%%%%%%%%%%%%%%%%%%%%%%%%%%%%%%%%%%%%%%%%%%%
%%%%%%%%%%%%%%%%%%%%%%%%%%%%%%%%%%%%%%%%%%%%
%%%%%%%%%%%%%%%%%%%%%%%%%%%%%%%%%%%%%%%%%%%%
%%%%%%%%%%%%%%%%%%%%%%%%%%%%%%%%%%%%%%%%%%%%
% Se aclara que el producto de distribuciones tiene sentido con la formula de convolucion LOCALMENTE

In this case, the product is calculated \textit{locally} using the convolution formula in the following sense \cite{brouder2014smooth}. For every $x\in\mathbb{R}^n$ there exists an $f\in\mathcal{D}$ such that $f\equiv 1$ in a neighborhood of $x$ such that the following integral is absolutely convergent

\begin{equation}
    \int \frac{dq}{(2\pi)^n}\widehat{f\Gamma}_1(k-q)\widehat{f\Gamma}_2(q),\ \ \ \ \ \forall k\in\mathbb{R}^n
\end{equation}

%%%%%%%%%%%%%%%%%%%%%%%%%%%%%%%%%%%%%%%%%%%%
%%%%%%%%%%%%%%%%%%%%%%%%%%%%%%%%%%%%%%%%%%%%
%%%%%%%%%%%%%%%%%%%%%%%%%%%%%%%%%%%%%%%%%%%%
%%%%%%%%%%%%%%%%%%%%%%%%%%%%%%%%%%%%%%%%%%%%

Some useful properties of the wavefront set are presented here. For $f,\ g$ distributions and $F=f\otimes g$ the tensor product, in the sense $F(x,y)=f(x)g(y)$; then

%%%%%%%%%%%%%%%%%%%%%%%%%%%%%%%%%%%%%%%%%%%%
%%%%%%%%%%%%%%%%%%%%%%%%%%%%%%%%%%%%%%%%%%%%
%%%%%%%%%%%%%%%%%%%%%%%%%%%%%%%%%%%%%%%%%%%%
%%%%%%%%%%%%%%%%%%%%%%%%%%%%%%%%%%%%%%%%%%%%
% Se corrige la propiedad del WF set del producto tensorial

\begin{equation}
    WF(F)\subseteq WF(f)\times WF(g)\cup\left((supp\ u\times\{0\})\times WF(g)\right)\cup\left(WF(f)\times(supp\ v\times\{0\})\right),
\end{equation}

%%%%%%%%%%%%%%%%%%%%%%%%%%%%%%%%%%%%%%%%%%%%
%%%%%%%%%%%%%%%%%%%%%%%%%%%%%%%%%%%%%%%%%%%%
%%%%%%%%%%%%%%%%%%%%%%%%%%%%%%%%%%%%%%%%%%%%
%%%%%%%%%%%%%%%%%%%%%%%%%%%%%%%%%%%%%%%%%%%% 

in particular, if the distribution is independent of some set of variables $G(x,y)=g(x)$

\begin{equation}
    WF(G)=\left\{(x,y;k_1,k_2)|(x;k_1)\in WF(g)\right\}.
\end{equation}

From translational invariance, a distribution $F(\underline{x},x)=f(\underline{x}-x)$ has a wavefront set

\begin{equation}
    WF(F)=\left\{(\underline{x},x;\underline{k},k)|(\underline{x}-x;\underline{k})\in WF(f),\ \sum_{j=1}^{n}k_j+k=0\right\}.
\end{equation}

Furthermore, the following wavefront sets for the delta, step and Wightman two-point distributions turn to be useful for our problem \cite{brouder2014smooth}

\begin{align}
    WF(\delta(x))&=\{(x;k)|x=0,\ k\neq 0\}\\
    WF(\theta(x))&=\{(x;k)|x=0,\ k\neq 0\}\\
    WF(\Delta_m^+(x))&=\{(x;k)|k_0=|\vec{k}|,\ x^0=\lambda k_0,\ \vec{x}=-\lambda \vec{k},\ \lambda\in\mathbb{R}\}.
\end{align}

Finally, for the definitions \eqref{eqlambda1} and \eqref{eqlambda2} for the Fourier transform of the non-local kernel $\Gamma_n$ which defines the interaction \eqref{inter}, it is possible to extract information about the structure of the singularities. Let us define the algebraic projective varieties

\begin{equation}
    K_{\pm}=\left\{(k^1,\ldots,k^n)\in\mathbb{R}^{4n}|\sum_{1\leq j<l\leq n}\left(k_0^jk_i^l-k_0^lk_i^j\pm \epsilon^{imn}k^j_mk^l_n\right)=0,\ i=1,2,3\right\},
\end{equation}

both of codimension 3 in $\mathbb{R}^{4n}$. The Fourier transform $\Lambda_n$ in \eqref{eqlambda1} is a non-zero constant for any point in $K_0:=K_+\cup K_-$ and decays at least as $k^{-2}$ for any other direction outside $K_0$. These later directions correspond to a behavior of a continuous function, and can be avoided, for example by

\begin{equation}
    \begin{split}
        \Lambda_n&=\Lambda_n^{(c)}+\Lambda_n^{(\delta)}\\
        \Lambda_n^{(\delta)}&(\underline{k})=e^{-\beta_+^2-\beta_-^2}\\
        \Lambda_n^{(c)}&=\Lambda_n-\Lambda_n^{(\delta)}.
    \end{split}
\end{equation}

$\Lambda_n^{(\delta)}$ has a $\delta$-like singularity at directions in Fourier space in $K_0$. The distribution $\Gamma^{(\delta)}_n$, whose Fourier transform is $\Lambda_n^{(\delta)}$, has a wavefront set

\begin{equation}
    WF(\Gamma_n^{(\delta)})\subseteq\left\{(\underline{x};\underline{k})|(\underline{x})\in K_0,\ (\underline{x})-\lambda(\underline{k})\in K_0\ \forall \lambda\in\mathbb{R}\right\}
\end{equation}

Once the wavefront sets of the involved distributions are described, standard renormalization can be performed, the product of distributions of the form $\theta(\Delta^+_m)^k\Gamma_n$ as a distribution in $\mathbb{R}^{4(n+1)}$ is not well-defined everywhere. Let us call $\Omega$ the subset where the H\"ormander criterion for product of distributions does not hold; then the product is a well defined distibution in $\mathcal{D}'(\mathbb{R}^{4(n+1)}\backslash\Omega)$. The renormalization should be again a problem of extension of distributions to $\mathcal{D}'(\mathbb{R}^{4(n+1)})$ satisfying a set of renormalization axioms \cite{dutsch2019classical}. The main concern here should be the scaling and mass expansion which would require an expansion in power series in $\lambda_P$ which should scale appropiately. This procedure is beyond the scope of the present work.

\section{Conclusion and perspectives}
\label{sec:7}

The $\phi^n$ interacting scalar field presented in \cite{doplicher1995quantum} has been described in the framework of pAQFT as a non-local effective theory and the set of modified Feynman rules has been obtained. The effects of noncommutativity of the spacetime are encoded in the non-local distributions $\Gamma_n$. The structure of their singularities has been described 
using techniques of microlocal analysis.
Each distribution $\Gamma_n$ is expressed as the sum of a continuous function plus a term that carries a $\delta$-type singularity, whose wavefront set can be described explicitly, up to the solution of a system of second order algebraic equations (geometrically given by a projective variety).\\

In contrast to the result of \cite{doplicher2020perturbative}, where the effective non-local QFT turns up to be finite at any order in perturbation thery; our approach to pAQFT based on the original work of Doplicher et. al. \cite{doplicher1995quantum} keeps some non-trivial singularities which have to be treated using some renormalization scheme (e.g. BPHZ or Epstein-Glaser).
These issues will be explored in a forthcoming publication. A treatment at finite temperature or even a generalization 
to curved backgrounds appears to be accessible in the presented framework.

\section*{Acknowledgments}
  A.F. Reyes-Lega acknowledges financial support from the
Faculty of Sciences of Universidad de los Andes through
project No. INV-2021-128-2324. J. F. L\'opez acknowledges financial support from the
Faculty of Sciences of Universidad de los Andes through
projects  No. INV-2022-144-2493 and No. INV-2022-154-2638.

%\bibliography{sample}


\end{document}